\pdfoutput=1
\documentclass[12pt]{article}

\usepackage[utf8]{inputenc}
\usepackage[T1]{fontenc}
\usepackage{geometry}
\usepackage{setspace}
\usepackage{authblk}
\usepackage{amsmath,amssymb}
\usepackage{graphicx}
\usepackage{booktabs}
\usepackage{array}
\usepackage[round]{natbib}
\usepackage{csquotes}
\usepackage{tikz}
\usetikzlibrary{positioning,arrows.meta,calc}
\usepackage[hidelinks]{hyperref}
\usepackage{xcolor}

\newcommand{\keywords}[1]{\par\noindent\textbf{Keywords:}\ #1}
\newcommand{\na}{--}

\title{\textbf{Measuring the Inaccessible:\\ Can AI Evaluate Accessibility in Heritage Tourism?}}

\author[1]{Sabrina Armenio\thanks{ORCID: 0000-0002-8386-2938.}}
\author[2]{Manuele Leonelli\thanks{Corresponding author. School of Science and
Technology, IE University, IE Tower, Paseo de la Castellana 259E, 28046 Madrid, Spain.
Email: manuele.leonelli@ie.edu. ORCID: 0000-0002-2562-5192.}}
\affil[1]{Departamento de Educaci{\'o}n y Patrimonio, Universidad Aut{\'o}noma de Madrid, Madrid, Spain}
\affil[2]{School of Science and Technology, IE University, Madrid, Spain}
\date{}

\hypersetup{pdftitle={Measuring the Inaccessible: Can AI Evaluate Accessibility in Heritage Tourism?},
  pdfauthor={Sabrina Armenio, Manuele Leonelli}}
\begin{document}
\maketitle

\begin{abstract}
\noindent
Accessibility assessment underpins the right to participate in heritage tourism, yet evaluation of archaeological sites remains artisanal: fieldwork does not scale, is often carried out without accessibility training, and is reported in forms that reach neither the sector nor disabled visitors. AI appears to offer a shortcut. We argue that assessment is measurement and inference under sparse, expert-mediated evidence: AI transforms how that evidence is captured and communicated, whereas evaluation, which turns evidence into defensible judgements, demands properties current generative systems lack. In a five-park field study in Spain and Italy, an AI desk audit scored only 66\% of dimension-site combinations, could not observe embodied and organisational dimensions, and mirrored sites' online self-presentation. Simple Bayesian models instead quantify the confidence rankings deserve and calibrate desk evidence against fieldwork. We propose a separation of roles between AI and interpretable statistical evaluation as the basis for a reliable accessibility atlas.
\end{abstract}


\section{Introduction}
\label{sec:intro}

Cultural heritage, from archaeological sites and museums to intangible assets, shapes
collective memory, identity, and knowledge, and access to it is internationally
recognised as a right: the \citet{UNCRPD2006} affirms participation in
cultural life on an equal basis, the \citet{Faro2005} recognises everyone's right to
engage with cultural heritage, the \citet{EAA2019} sets binding accessibility
requirements across the European Union, and \citet{UNESCO1972} and \citet{ICOMOS2008}
frame inclusive interpretation as integral to responsible heritage management. The
population concerned is far from marginal: an estimated 1.3 billion people, 16\% of the
world's population, live with a significant disability \citep{WHO2022}, and on a
whole-of-life view access requirements touch an estimated one in three people through
impairment, ageing, or travelling with others \citep{DarcyDickson2009,BuhalisDarcy2011}.
Yet a persistent gap separates these principles from lived experience. Policy and
practice have traditionally concentrated on museums and urban institutions, whereas
archaeological parks present physical, sensory, cognitive, digital, cultural, and
economic barriers of their own \citep{Nicolini2021}, so that for many visitors they
remain effectively closed by design rather than intent. Inclusive archaeology requires
more than physical access: because engagement with heritage is embodied, sensory, and
cognitive \citep{Armenio2026}, accessibility cannot be confined to the visual dimension,
and management depends on understanding how the barriers interact to shape the visit.

The deeper obstacle is that the problem cannot yet be observed at scale. Rigorous
assessment remains artisanal, relying on expert fieldwork that is time-consuming, costly,
and difficult to replicate, and routine practice compounds the difficulty: assessments
are frequently carried out by heritage professionals without specific training in
accessibility, and their results are communicated in forms that serve neither the wider
sector nor the disabled visitors whose decisions depend on them \citep{Nigg2025}. The
field study underpinning this paper \citep{Armenio2026} evaluated five parks across
Spain and Italy over several years, while managers, policymakers, and disabled travellers
increasingly require comparable information on a far larger scale. That challenge has
emerged alongside the rapid uptake of AI in tourism
\citep{Gursoy2023,CarvalhoIvanov2024,Sigala2024,HallCooper2025}, where web-scale text,
imagery, and review data appear to promise accessibility assessment without fieldwork.
The critical questions are what kind of evidence AI-produced accessibility information
represents, how reliable it is, and whether it can be compared meaningfully with expert
field-based assessment.

This paper argues that accessibility assessment is, at root, probabilistic measurement
and inference under sparse, heterogeneous, expert-mediated evidence. AI has
transformative potential for the \emph{capture} and \emph{communication} of
accessibility information. The \emph{evaluative} step, which turns imperfect evidence
into comparable, defensible judgements, demands interpretability, calibrated uncertainty,
and accountability to the legal frameworks that underpin the right to cultural
participation, properties current generative systems cannot reliably provide on their
own. We therefore propose an explicit separation of roles between AI-based capture and
communication and statistical evaluation, and provide pilot evidence for it. The
contribution is threefold. Conceptually, we develop the distinction between communicating
accessibility and evaluating it, and set out the epistemic limits of automated evaluation
(Section~\ref{sec:framework}). Empirically, we report what we believe to be the first
direct comparison of an AI desk audit with an expert fieldwork audit of the same heritage
destinations. Methodologically, we show how simple Bayesian hierarchical models turn
audit scores into judgements that carry their own uncertainty and calibrate desk evidence
against fieldwork (Section~\ref{sec:casestudy}), before proposing an interpretable
framework and discussing its implications (Sections~\ref{sec:method} and
\ref{sec:agenda}).

\section{Conceptual framework: accessibility assessment as measurement under uncertainty}
\label{sec:framework}

\subsection{Accessibility as a multidimensional, latent construct}
\label{sec:measurement}

Accessibility is not a single, observable property of a heritage site but a
multidimensional \emph{latent} construct inferred from indirect evidence, and it is
embodied and interpretive as much as infrastructural. Visitors encounter it through
movement, orientation, sensory perception, fatigue, and interpretation, so the same
pathway is experienced differently according to age, physical or cognitive profile,
sensory abilities, weather, or group composition. Assessment therefore concerns the
situated experience of navigating and understanding a site, not only the presence of
accessible features: a visitor may physically reach a site while remaining excluded
from its interpretive content.

The field study on which this paper draws \citep{Armenio2026} operationalises it
through seven dimensions: physical, sensory, and cognitive accessibility; inclusive
programming; staff training; technology and digital provision; and evaluation and
monitoring. Each is evidenced by heterogeneous indicators: binary infrastructure
checklists, Likert-scale visitor responses, field observation, and documentary
analysis. These seven dimensions are the instrument used throughout.

Indicator-based assessment of heritage destinations has a respectable tradition in
tourism research \citep{McKercherHo2006,NaniopoulosTsalis2015}, and the most direct
precedent for archaeological parks is the Heritage Site Accessibility Tool of
\citet{Salva2023}: twenty-nine indicators in two weighted areas, applied through
fieldwork to eight Mediterranean World Heritage sites, including one studied here. Its
value lies in the author's explicit acknowledgement of its limits: subjective weights, a
summary percentage that ``can be seen as questionable'', a single person's judgement,
and an access chain only as strong as its weakest link. These are the well-documented
pathologies of composite indicators \citep{OECD2008,Greco2019}: subjective weighting,
information loss under aggregation, and false precision. This paper asks what assessment
should look like once those limitations are built into the measurement process, at a
time when AI makes accessibility scores cheap to generate at scale.

Three features of the construct govern the approach taken here. First, accessibility is
latent and assessment is assessor-mediated: a published score is a potentially noisy
reading produced by a particular assessor from particular evidence, not a direct
measurement of the site. Second, the dimensions are interdependent: terrain constraints
affect sensory experience, orientation, cognitive accessibility, and the interpretation
feasible on site. Third, the evidence is sparse and heterogeneous: few sites, many
indicators, missing observations, and often a single assessor per site. This regime
determines which analytical tools are appropriate, and it is where naive automation is
most misleading. The challenge is not primarily technological but epistemological: which
aspects of accessibility can be represented as data, how heterogeneous evidence should
be combined, and which dimensions remain dependent on situated human judgement.

\subsection{Communication versus evaluation: two roles of AI}
\label{sec:promise}

The contributions AI can make to accessible heritage tourism divide into two roles that
are routinely conflated and make fundamentally different epistemic demands:
\emph{communicating} accessibility and \emph{evaluating} it.

The communication role is where the potential of AI is already substantial, and where
current practice is visibly deficient: the accessibility information fieldwork produces
is typically reported in technical or administrative formats that never reach the
visitors whose decisions depend on it. Information has long been the critical resource
of the accessible tourism market \citep{MichopoulouBuhalis2013}, and its technological
provision has evolved steadily \citep{NiggPeters2022}. Large language and multimodal
models now generate easy-read text, multilingual audio description, and adaptive
wayfinding support at very low marginal cost, and domain-tuned models can adapt these
outputs to specific audiences \citep{ChemnadOthman2024,Tuomi2025}; AI here operates as
an accessibility \emph{intervention} in its own right. Closely related is the capture of
evidence: multimodal models can structure the scattered, multilingual documentation on
which assessment depends, computer vision can extract preliminary cues about terrain and
barriers from imagery, and visitor reviews can reveal signals of lived experience that
fieldwork rarely captures at scale \citep{CarvalhoIvanov2024,Sigala2024}. By lowering
the marginal cost of capture and communication, these technologies make a broad,
multi-site atlas conceivable at all, although digital channels can exclude older or less
connected visitors, so they should complement rather than replace non-digital provision
\citep{Gursoy2023}.

Evaluation is a different act: turning evidence into a judgement that scores, compares,
prioritises, or certifies sites in ways that affect decisions and should remain open to
scrutiny. A communicated description can be questioned by the reader; an evaluative score
travels into rankings, funding decisions, and compliance assessments, often while the
provenance and uncertainty of the evidence remain invisible. The distinction matters
especially in heritage accessibility, where evaluation concerns a latent construct
inferred from sparse, expert-mediated evidence: an AI system can efficiently identify and
organise relevant information without establishing the validity of the judgement derived
from it. The evaluative role demands interpretability, calibrated uncertainty, and
accountability, which current generative systems cannot guarantee as autonomous
evaluators.

The position advanced here is therefore in favour of \emph{appropriately placed} AI:
transformative in communication and capture, out of place at the evaluative core, where
it should be overseen by transparent statistical methods and accountable human judgement.
This is consistent with work on hybrid and augmented intelligence in tourism
\citep{Tuomi2025}, and with the argument of \citet{Leonelli2026glassbox} that where
opacity is institutionally and legally untenable, generative systems should be mediated
by transparent probabilistic structures rather than explained after the fact. AI is
accordingly one component of a hybrid evidence architecture, expanding capture and
communication, while interpretable inference provides the evaluative layer.

\subsection{Data sparsity, embodiment, and measurement validity}
\label{sec:validity}

The data regime of heritage accessibility (tens of sites, dozens of indicators, missing
entries, and frequently one assessor per site) is where data-hungry, black-box approaches
are weakest and explicit statistical models, which state their assumptions and can
incorporate expert knowledge, are strongest. Heritage accessibility differs further from
that of contemporary built environments because interventions are constrained by
conservation and authenticity: archaeological parks occupy uneven terrain, contain
fragile remains, and sit under protection regimes that limit physical modification, so
assessment must weigh conservation, authenticity, and inclusion together, recognising
that some barriers are structural features of the resource rather than failures of
management \citep{Armenio2026}.

The question of who, or what, does the scoring then runs deeper than data quality. An
automated system scoring a site from its website measures the digital representation of
the site, not the site itself; desk-based and fieldwork-based scores are different forms
of evidence that should not be treated as interchangeable without explicit calibration.
Field observation across the five parks repeatedly revealed discrepancies between
documented and experienced accessibility: official materials did not consistently capture
terrain, gradients, wayfinding difficulties, sensory demands, or the cumulative effects
of fatigue \citep{Armenio2026}. Accessibility is in part embodied and experiential,
realised in the encounter between a particular visitor and a particular place
\citep{BuhalisDarcy2011}: documentation records that a provision exists, not whether it
works for a visitor negotiating uneven terrain or managing fatigue. Multisensory
engagement is part of what makes heritage visits meaningful \citep{Moscardo1996}, and
much of it is invisible from a desk. AI-assisted documentary assessment should therefore
not be assumed to reproduce fieldwork more cheaply, since it may be a different
measurement process with its own sources of error. Section~\ref{sec:casestudy} makes the
gap measurable.

\subsection{False accessibility claims and the cost of unreliable evidence}
\label{sec:falseclaims}

The failure mode with the most direct human cost is the confident, wrong accessibility
claim. Generative systems produce fluent assertions not necessarily anchored to verified
evidence; accuracy is among the most consistently flagged risks of their deployment in
tourism \citep{Gursoy2023,CarvalhoIvanov2024}, and inaccurate generative output
measurably distorts travel decisions \citep{Christensen2025,Kim2025}. The costs of error
are sharply asymmetric. A site wrongly described as inaccessible loses a visitor; a
visitor who travels on the strength of an inaccurate claim about a step-free route, an
adapted toilet, or current opening arrangements pays in money, fatigue, and dignity.
These consequences fall hardest on disabled travellers, who plan in advance and depend on
accurate information to judge whether a visit is feasible at all
\citep{DarcyDickson2009,MichopoulouBuhalis2013,Salva2023}.

\citet{Salva2023} observes that even the blanket label ``accessible'' misleads when
detached from the specific conditions under which accessibility is experienced; a
system that synthesises such labels at scale may amplify existing weaknesses in
accessibility communication rather than resolve them. What matters is not only whether
an individual statement is correct, but whether users can rely on it and understand its
evidential basis.

Emerging legal practice reflects the same risk. In \citet{Moffatt2024} a tribunal held
an airline liable for negligent misrepresentation after its chatbot gave a customer
inaccurate fare information, finding that the company could not disclaim responsibility
for what its automated system asserted. The case concerned fares rather than
accessibility, but the principle carries: using an automated system does not remove
responsibility for the claims users are encouraged to rely upon. Reliability is therefore
a prerequisite for any assessment meant to inform decisions, comparisons, or policy, and
where AI outputs become evaluative claims their evidential basis, uncertainty, and
provenance must remain visible.

\subsection{Interpretability, accountability, and the limits of aggregation}
\label{sec:aggregation}

When a score informs funding allocation, management priorities, or a compliance
assessment under the \citet{EAA2019} or \citet{UNCRPD2006}, ``the model returned 2 out
of 5'' is not an adequate basis for a decision. Such decisions require a traceable
account of how the score was derived, the evidence, the assumptions, and the uncertainty,
which opaque scoring cannot supply. Tourism research has flagged the legal and compliance
implications of integrating generative systems with heritage information as an open
question \citep{Sigala2024}; this paper treats the evaluative layer as central rather
than as a technical afterthought.

A single opaque number fails accountability in a second way: by aggregating away the
structure a decision needs. On the visitor side, needs are heterogeneous: measures that
improve access for wheelchair users may offer little to visitors with sensory
impairments, autism spectrum conditions, cognitive disabilities, or age-related
limitations, so a single score says little about for whom a site is accessible and where
barriers remain. On the site side, a mean across dimensions obscures the mechanisms
managers must act on: terrain constraints propagate into orientation, sensory experience,
fatigue, and cognitive accessibility, whereas digital or interpretive gaps are
comparatively easy to address, so two parks can share an average while facing different
management challenges. Tourism's own instruments show it. Reviewing the market-appeal and
robusticity matrix of \citet{duCros2001}, \citet{McKercherHo2006} note that composite
scores cluster near the midpoint as sub-dimensions wash out, and that equal weighting
denies a decisive flaw the weight it deserves. \citet{Salva2023} makes the same point in
the language of the access chain: one broken link can void the visit for particular
audiences, however good the average.

The objective is not to eliminate summary scores but to place them within an
interpretable structure that shows which dimensions drive an assessment, what evidence
supports each judgement, where information is missing, and how confident the comparison
is, preserving the value of a summary while avoiding the false precision of a single
unexplained number.

\section{Case study: fieldwork versus AI desk audits of five archaeological parks}
\label{sec:casestudy}

The claims above are testable. We use the five archaeological parks assessed in situ
between 2022 and 2026 by \citet{Armenio2026}: Cueva Pintada (Gran Canaria, Spain), the
Valle dei Templi (Agrigento, Italy), Cerro de la Cabeza (Valdepe\~nas, Spain), Tiermes
(Soria, Spain), and Solunto (Sicily, Italy). Together they span the realistic range of
the sector, from a purpose-built museum-park to minimally staffed open-air sites, and so
allow a direct test of whether online accessibility information corresponds to
accessibility as experienced in the field.

\subsection{The fieldwork benchmark}

Table~\ref{tab:audit} (upper row of each pair) reports the field assessment of each
park on the seven dimensions, on a 1 to 5 scale \citep{Armenio2026}. The benchmark
combines binary indicator grids and structured observation under a common protocol at
every park, complemented by visitor questionnaires and documentary analysis where
available, with each score anchored to explicit descriptors rather than to an overall
impression. The same assessor, trained specifically in accessibility assessment, applied
the same instrument across all five parks, so between-park comparisons are internally
consistent, and the benchmark is deliberately stronger than the routine practice
described in Section~\ref{sec:intro}. Its own uncertainty is unquantified, however: with
one assessor, assessor effects cannot be separated from the site and dimension
components, and the noise term of the model in Section~\ref{sec:uncertainty} captures
within-assessor inconsistency only.

Two observations matter for what follows. First, no park dominates: Cueva Pintada has
the highest overall profile but is not strongest in every dimension, and the weakest
dimension differs across parks, which is the structure a single average obscures. Second,
evaluation and monitoring is a systematic weakness, at or near the bottom for four of the
five parks: provision is more often implemented than evaluated.

\begin{table}[t]
\centering
\caption{Field assessment versus AI desk audit of the five parks across the seven
dimensions (1 to 5 scale). Upper rows (F): fieldwork scores of \citet{Armenio2026};
lower rows (D): first run of the pilot AI desk audit (6 July 2026), scoring only where
retrieved online evidence supported a judgement (--\ = insufficient evidence).}
\label{tab:audit}
\small
\begin{tabular}{llccccccc}
\toprule
Park & & Phys. & Sens. & Cogn. & Progr. & Staff & Tech. & Eval. \\
\midrule
Cueva Pintada        & F & 4 & 4 & 3 & 4 & 3 & 4 & 3 \\
                     & D & 5 & 4 & 4 & 4 & 3 & 4 & \na \\
\addlinespace
Valle dei Templi     & F & 4 & 4 & 2 & 3 & 3 & 4 & 1 \\
                     & D & 4 & 4 & 2 & 3 & \na & 4 & 3 \\
\addlinespace
Cerro de la Cabeza   & F & 2 & 3 & 2 & 3 & 2 & 2 & 2 \\
                     & D & 2 & \na & 3 & 3 & \na & 4 & \na \\
\addlinespace
Tiermes              & F & 3 & 2 & 2 & 3 & 2 & 2 & 1 \\
                     & D & 2 & \na & 3 & 2 & \na & 1 & 2 \\
\addlinespace
Solunto              & F & 3 & 2 & 2 & 2 & 2 & 2 & 1 \\
                     & D & 2 & \na & \na & \na & \na & 3 & \na \\
\bottomrule
\end{tabular}
\end{table}

\subsection{The pilot desk audit: design and findings}
\label{sec:pilot}

We therefore ran a pilot desk audit in which an AI system assessed each park on the same
seven dimensions, from publicly available online information only, under a fixed
protocol: score a dimension only when retrieved evidence supports a judgement, otherwise
record it as unassessable, and log the evidence behind every score. The first run, in
Table~\ref{tab:audit}, was performed on 6 July 2026, in a single pass, by Claude Fable 5
(Anthropic) with extended reasoning and integrated web search; \ref{sm:protocol} gives the verbatim
prompt and \ref{sm:log} the evidence log, and two replication runs follow in
Section~\ref{sec:fieldnotes}. The pilot is illustrative evidence of the measurement
problem, not a benchmark of any particular system.

Three findings stand out. First, coverage rather than accuracy was the binding
constraint. The model could justify a score for 23 of 35 dimension-park combinations
(66\%), and the missingness is clearly non-random: physical and technological provision
could be scored for every park, whereas staff training was unassessable for four parks
and sensory provision and evaluation for three each. The desk saw the dimensions most
visible through websites and far less of the embodied and organisational dimensions that
Section~\ref{sec:measurement} identified as constitutive. Coverage also tracked digital
prominence rather than accessibility: the two flagship destinations were almost fully
assessable, whereas Solunto, the least documented park, yielded evidence for two of seven
dimensions. An automated atlas built this way would inherit, and could amplify, that
visibility gap.

Second, where the desk and the field disagreed, the disagreement was systematic. Across
the 23 scoreable combinations the mean absolute deviation from the fieldwork was 0.61
points, and the direction of several discrepancies is informative. The desk overscored
where official self-description or announced provision dominated the record: it gave
Cueva Pintada the maximum for physical access on the strength of the site's own ``fully
accessible'' materials where the field recorded 4, credited the Valle dei Templi with
active evaluation (3 against a field score of 1) on the basis of publicised participatory
initiatives, and rated the technology of Cerro de la Cabeza 4 against a field score of 2,
reflecting announced virtual-reality provision that the fieldwork found less effective in
practice. Conversely, it could not score provision that was present but poorly
documented, as with the same park's sensory resources, observed in the field yet absent
from the web. On this evidence, the desk audit is a different instrument from the field audit,
measuring a different construct, principally a site's digital self-presentation, so its
limitations arise from a difference in construct as well as from prediction error.

Third, the instruments may disagree simply because they observe at different times: the
desk captures what was online on the audit date, fieldwork the site as visited, so any
atlas combining them must carry time stamps as well as provenance. None of this makes
desk assessment useless: where provision is adequately documented, agreement with the
field was close, which is why the capture layer of Section~\ref{sec:promise} is
valuable. The question is whether documentary evidence can be calibrated against
fieldwork rather than whether it can replace it, and Section~\ref{sec:calibration} shows
that calibration is a modelling task the data can support.

\subsection{Qualitative observations from the evidence log}
\label{sec:fieldnotes}

The evidence log (\ref{sm:log}) shows why documentary coverage and field-based accessibility
diverge. Some information never appeared online: no park documented staff
disability-awareness training, the closest evidence being the generic presence of guides,
and systematic evaluation left traces only as announcements, publicity for an event or a
dated mention of a visitor survey, insufficient to distinguish a monitoring practice from
a one-off communication. Sensory provision at the smaller parks was undocumented even
where the fieldwork had found it. Other information appeared but could not be
adjudicated: for the Valle dei Templi the system retrieved both the official designation
of the site as fully accessible and independent reports of steep gradients and scarce
accessible parking, a contradiction a human auditor would investigate and a fluent
generative system could resolve without making the uncertainty visible. A further pattern
was temporal: for two parks the strongest online evidence concerned works announced after
the fieldwork period, so the desk partly assessed a site's anticipated condition. One
observation cuts the other way. No run produced fabricated provision, but this is not
evidence that generative AI is inherently reliable: the protocol required supporting
evidence and permitted leaving a dimension unassessed, so the restraint is attributable
to the procedure rather than the model, which is the design lesson
Section~\ref{sec:method} encodes.

Generative systems are stochastic and the web changes, so reproducibility was tested
directly: the audit was repeated on 18 September 2026 in ten fresh sessions, one per park
for each of two models (Claude Fable 5.1 and Claude Opus 5), under the same instruction
and without access to the field scores or the first-run log (\ref{sm:runs}). Coverage was 23, 23,
and 26 of 35; whether a cell was assessable agreed in 74 to 86 percent of cells between
pairs of runs, and six cells (staff training at three parks, evaluation at two, sensory
at one) were unassessable in every run. Where two runs both scored a cell they agreed
within one point in 90 to 100 percent of cases, and the mean absolute deviation from the
field was 0.61, 0.65, and 0.69. The three discrepancies above kept their direction in
every run, with smaller magnitude in September, while Cueva Pintada's sensory provision
fell to 2 against a field score of 4 because the park's own accessibility page could not
be retrieved that day: run-to-run variation arose as much from what the retrieval layer
could reach as from the model.

The limits of the pilot remain: five parks in two countries, one vendor, three
single-pass runs on two dates. The findings are evidence for the conceptual and
methodological argument, not a general benchmark of AI performance; a fuller evaluation
would add other vendors' models and more runs under a frozen prompt, with coverage,
inter-run agreement, deviation from the field, and the rate of unsupported claims as
pre-registered metrics.

\subsection{A Bayesian hierarchical model of the fieldwork scores}
\label{sec:uncertainty}

The desk audit shows what happens when evidence collection is mistaken for evaluation;
the constructive question is what evaluation should look like once assessment is
recognised as probabilistic inference from imperfect evidence. Bayesian models suit this
problem because they make uncertainty, structure, and prior information explicit objects
of inference that others can inspect and contest, rather than by-products of aggregation
\citep{gelman1995bayesian}. We illustrate with the simplest useful instance, applied to
the fieldwork data.

Each field score is decomposed additively into an overall level, a site component, and a
dimension component, all estimated with full posterior uncertainty; the specification,
priors, and computation are given in \ref{sm:models}. This is the least structured model sufficient to
demonstrate uncertainty propagation; ordinal likelihoods, latent-variable formulations,
and explicit dependence between dimensions belong to the validated instrument of
Section~\ref{sec:agenda}. Two caveats follow. With thirty-five observations the exercise
illustrates the reasoning rather than establishing findings about the parks. A site-level
summary may also seem to reproduce the aggregation problem of
Section~\ref{sec:aggregation}; the choice is deliberate, because league tables publish
such one-number summaries, and the point is to show how one can carry explicit
uncertainty and dimension-level information.

\begin{figure}[t]
\centering
\includegraphics[width=.92\linewidth]{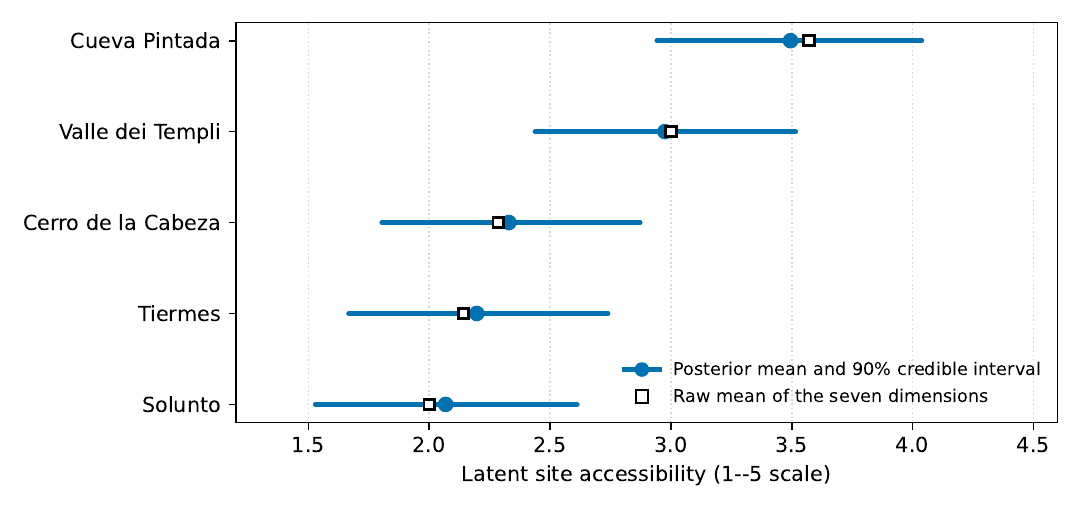}
\caption{Latent site accessibility for the five fieldworked parks: posterior means with
90\% credible intervals (filled circles with horizontal bars), against the raw mean of
the seven dimensions (open squares). The intervals of the three lowest-scoring parks
overlap almost entirely: the ordering implied by their raw averages holds with posterior
probability 0.41.}
\label{fig:posterior}
\end{figure}

Figure~\ref{fig:posterior} reports the site-level result. The raw averages produce an
apparently precise ordering: 3.57, 3.00, 2.29, 2.14, 2.00. The posterior supports a more
cautious reading. Cueva Pintada is credibly ahead of the Valle dei Templi (posterior
probability 0.96), and both are clearly separated from the remaining three, but the
bottom three parks are statistically indistinguishable: the probability that Cerro de la
Cabeza exceeds Tiermes is 0.68, that Tiermes exceeds Solunto 0.68, and the probability
that the raw ordering of all three is correct is 0.41, barely better than chance.
Treating those decimals as real differences would claim a certainty the evidence does not
support. The dimension components carry information of their own: evaluation and
monitoring is credibly the weakest dimension, with posterior effect $-0.87$ (90\%
credible interval $-1.45$ to $-0.32$) and a 0.92 posterior probability of being the
lowest of the seven, confirming the exploratory observation above.

\subsection{Calibrating desk evidence against fieldwork}
\label{sec:calibration}

Sections~\ref{sec:promise} and \ref{sec:validity} argued that desk and field audits
measure different constructs; the statistical question is whether the relationship
between them can be learned. If it can, documentary evidence becomes a calibratable
complement rather than a rival, supplying the scalability an atlas requires while
fieldwork remains the reference. We therefore extend the model so that both audits are
generated by the same underlying site and dimension quantities, while desk scores are
allowed a dimension-specific systematic shift and their own noise level (\ref{sm:models}), drawn as a
graph in Figure~\ref{fig:model}. Estimating the shift is what calibration means.

The results, again illustrative, are consistent with the qualitative findings. The
estimated desk shift is centred above zero where announcement can substitute for evidence
of implementation: evaluation and monitoring (posterior mean $+0.5$, posterior
probability 0.89 of a positive shift), cognitive provision ($+0.5$, probability 0.89),
and technology ($+0.4$), and close to zero for physical provision, programmes, and staff.
The desk also remains the noisier instrument after correction, with residual standard
deviation 0.77 against 0.56 for the field. Combining the two sources narrows the
site-level intervals of Figure~\ref{fig:posterior} by roughly ten percent, a gain that
comes from estimating the relationship between the instruments rather than from taking
desk scores at face value. One limitation motivates future work: the desk's blanks are
informative, falling on the dimensions and parks least visible online, so a full atlas
must model the missingness rather than condition on it. Conditioning assumes a cell is
blank at random given site and dimension, which the coverage pattern contradicts, so the
calibration estimates should be read as conditional on online visibility.

\section{An interpretable framework for accessibility assessment}
\label{sec:method}

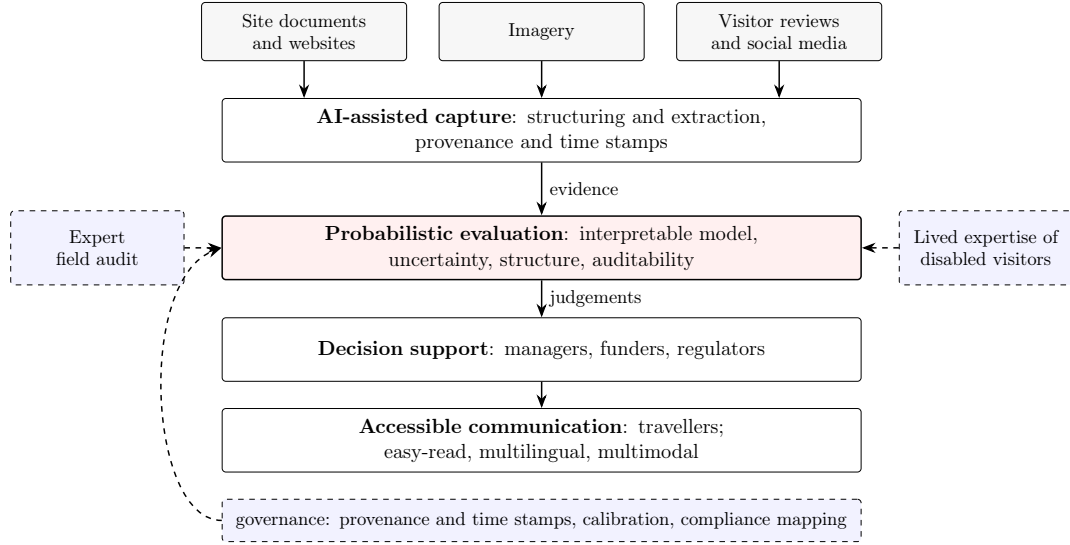
\begin{figure}
\centering
\scalebox{0.7}{
\begin{tikzpicture}[
  node distance=5mm and 6mm,
  src/.style={draw, rounded corners=2pt, align=center, minimum height=11mm,
              text width=36mm, font=\footnotesize, fill=gray!6},
  stage/.style={draw, rounded corners=2pt, align=center, minimum height=12mm,
                text width=118mm, font=\small},
  core/.style={stage, thick, fill=red!6},
  sidebox/.style={draw, dashed, rounded corners=2pt, align=center, font=\footnotesize,
                 text width=30mm, minimum height=14mm, fill=blue!5},
  band/.style={draw, dashed, rounded corners=2pt, align=center, font=\footnotesize,
               text width=118mm, minimum height=8mm, fill=blue!5},
  arr/.style={-{Stealth[length=2.6mm]}, thick},
  anchorarr/.style={-{Stealth[length=2.6mm]}, thick, dashed}
]
\node[src] (s1) {Site documents\\ and websites};
\node[src, right=of s1] (s2) {Imagery};
\node[src, right=of s2] (s3) {Visitor reviews\\ and social media};
\node[stage, below=7mm of s2] (cap) {\textbf{AI-assisted capture}: structuring and extraction,\\ provenance and time stamps};
\node[core, below=10mm of cap] (eval) {\textbf{Probabilistic evaluation}: interpretable model,\\ uncertainty, structure, auditability};
\node[sidebox, left=7mm of eval] (field) {Expert\\ field audit};
\node[sidebox, right=7mm of eval] (part) {Lived expertise of\\ disabled visitors};
\node[stage, below=7mm of eval] (dec) {\textbf{Decision support}: managers, funders, regulators};
\node[stage, below=of dec] (com) {\textbf{Accessible communication}: travellers;\\ easy-read, multilingual, multimodal};
\node[band, below=of com] (gov) {governance: provenance and time stamps, calibration, compliance mapping};
\draw[arr] (s1.south) -- (cap.north -| s1.south);
\draw[arr] (s2) -- (cap);
\draw[arr] (s3.south) -- (cap.north -| s3.south);
\draw[arr] (cap) -- node[right, font=\footnotesize]{evidence} (eval);
\draw[anchorarr] (field) -- (eval);
\draw[anchorarr] (part) -- (eval);
\draw[arr] (eval) -- node[right, font=\footnotesize]{judgements} (dec);
\draw[arr] (dec) -- (com);
\draw[anchorarr] (gov.west) .. controls +(-14mm,0) and +(-16mm,-6mm) .. (eval.west);
\end{tikzpicture}}
\caption{The proposed separation of roles for AI in heritage accessibility assessment:
AI-assisted capture structures heterogeneous evidence, an interpretable probabilistic
core turns it into judgements with uncertainty, AI-assisted layers support decisions and
communicate accessibly, and governance carries provenance, calibration, and compliance
throughout.}
\label{fig:framework}
\end{figure}

The case study motivates the separation of roles summarised in
Figure~\ref{fig:framework}. Heterogeneous evidence, from documents and imagery to visitor
reviews, is captured with AI assistance and handed to the evaluative core as
provenance-tagged, time-stamped material, never as a final judgement. The core is an
interpretable probabilistic model, anchored by expert fieldwork and the lived expertise
of disabled visitors, that turns evidence into judgements carrying uncertainty and an
auditable derivation, which then serve decision support for managers, funders, and
regulators, and accessible communication for travellers. AI sits at the edges;
accountable inference sits at the core.

\begin{figure}
\centering
\begin{tikzpicture}[
  lat/.style={draw, circle, align=center, minimum size=17mm, font=\footnotesize, fill=blue!4},
  obs/.style={draw, align=center, minimum size=13mm, font=\footnotesize, fill=gray!8},
  arr/.style={-{Stealth[length=2.4mm]}, thick},
  lbl/.style={font=\scriptsize, align=center}
]
\node[lat] (a) {site\\ level\\ $a_i$};
\node[lat, right=34mm of a] (b) {dimension\\ profile\\ $b_j$};
\node[lat, right=34mm of b] (d) {desk\\ distortion\\ $d_j$};
\node[obs, below=16mm of $(a)!0.5!(b)$] (F) {field score\\ $F_{ij}$};
\node[obs, below=16mm of $(b)!0.5!(d)$] (D) {desk score\\ $D_{ij}$};
\draw[arr] (a) -- (F); \draw[arr] (b) -- (F);
\draw[arr] (a) -- (D); \draw[arr] (b) -- (D); \draw[arr] (d) -- (D);
\node[lbl, below=1.5mm of F] {field noise $\sigma_F$};
\node[lbl, below=1.5mm of D] {desk noise $\sigma_D$ (larger)};
\end{tikzpicture}
\caption{The calibration model of Section~\ref{sec:calibration}. Circles are latent
quantities, squares observed scores, and arrows read ``is generated from'': field and
desk audits are noisy readings of the same underlying site and dimension quantities, with
the desk additionally subject to a dimension-specific distortion and larger noise. The
fuller programme adds arrows among dimensions and nodes for assessors and visitor
profiles.}
\label{fig:model}
\end{figure}
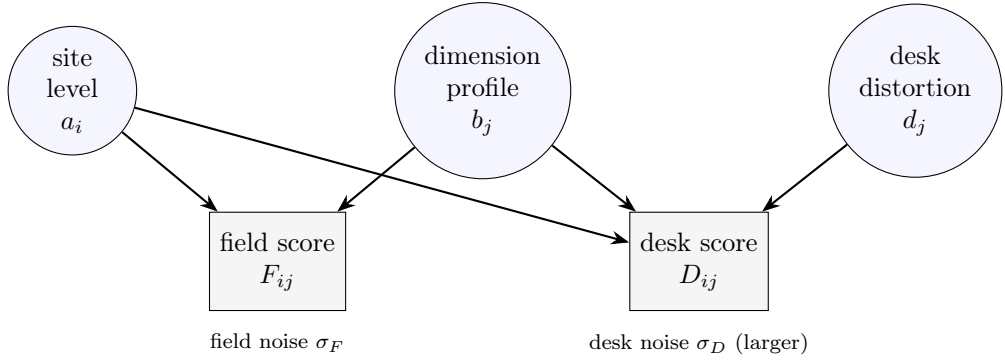

Figure~\ref{fig:model} draws that model as a graph that can be read aloud. Circles are
quantities nobody observes directly: each park's underlying accessibility level, each
dimension's profile across the sector, and the desk's dimension-specific distortion.
Squares are the recorded scores. The arrows say that both audits are generated by the
same underlying quantities, with the desk score additionally shifted by the distortion
and blurred by its own, larger noise. Every assumption is on the page: to contest an
assessment is to point at a node or an arrow. The same grammar extends directly, in the
spirit of glass-box architectures in which transparent probabilistic structure mediates
between model output and institutional decisions \citep{Leonelli2026glassbox}: arrows
among dimensions would encode how terrain constrains sensory and cognitive provision, and
further nodes would represent assessors or visitor profiles, in a tradition already
familiar in tourism \citep{HuangBian2009}.

Within this core, the answer to the limitations that \citet{Salva2023} acknowledges and
the composite-indicator literature formalises \citep{OECD2008,Greco2019} is direct: where
subjective weights are unavoidable, make them explicit objects of inference; where a
scalar conceals information, report the structure; where a point score overstates, report
the uncertainty. Faced with the same problem, \citet{McKercherHo2006} retreat to a
deliberately ordinal scheme; ours is the complementary move, keeping a graded assessment
but modelling what a flat score discards.

\section{Discussion: from pilot evidence to a living accessibility atlas}
\label{sec:agenda}

What stands between this pilot and the atlas the field needs is a set of connected
developments rather than a single missing technology or dataset.
Table~\ref{tab:programme} summarises what the case study showed and what would close each
gap. The rows are not independent: a validated instrument makes calibration meaningful,
calibration makes AI-assisted capture safe to use, and compliance-ready interpretability
lets any of it enter the decisions the \citet{EAA2019} and \citet{UNCRPD2006} regulate.

\begin{table}[t]
\centering
\caption{From pilot to atlas: what the case study showed and what is needed.}
\label{tab:programme}
\small
\begin{tabular}{p{0.24\linewidth}p{0.34\linewidth}p{0.34\linewidth}}
\toprule
Component & What the pilot showed & What is needed \\
\midrule
Validated instrument & Even expert scores support limited ranking confidence; assessor
effects not yet modelled & Multi-assessor studies; ordinal and latent-variable
formulations; explicit reliability \\
\addlinespace
AI-assisted capture & The desk cannot observe embodied and organisational dimensions,
is biased toward announcements, and varies with retrieval access & Multi-vendor,
repeated-run evaluation under frozen prompts; human validation; provenance and time
stamps \\
\addlinespace
Desk-field calibration & Dimension-specific desk distortion is estimable from paired
data (Section~\ref{sec:calibration}) & Joint models combining evidence streams;
treatment of informative missingness \\
\addlinespace
Compliance-ready interpretability & Opaque scores cannot answer regulator questions &
Glass-box models mapped to EAA and CRPD reporting duties \\
\addlinespace
Participation & Provision invisible online; lived experience absent from web data &
Co-design with disabled visitors across the whole assessment cycle \\
\bottomrule
\end{tabular}
\end{table}

The first four rows are statistical and infrastructural, and
Sections~\ref{sec:casestudy} and \ref{sec:method} have shown their simplest instances
working. Scaling them is also concrete. Under the calibration model the desk need not be
paired with fieldwork everywhere: a minimal atlas desk-audits every site, field-audits a
subsample stratified by digital prominence, since prominence drives missingness, and
estimates the desk distortions from the paired subset. Computation is linear in the
number of sites; what changes between five sites and five hundred is that assessor
effects become estimable once assessors overlap, missingness must be modelled rather than
conditioned on, and replicated desk runs become the bulk of the evidence.

The fifth row deserves more than a row. Accessible tourism is defined by collaborative
processes among stakeholders \citep{BuhalisDarcy2011}, and an instrument that scores the
experience of disabled visitors without their involvement perpetuates, in measurement,
the exclusion it purports to diagnose. Participation should extend beyond consultation to
the whole assessment cycle: indicator selection, field testing, interpretation of results,
and evaluation of interventions. This is established practice:
\citet{NaniopoulosTsalis2015} developed their monument-accessibility methodology through
a structured process that included people with disabilities directly, and
\citet{Salva2023} reaches the same conclusion for audit design. Co-design discharges an
ethical obligation and earns a methodological return at once, because lived expertise
identifies which indicators matter, which barriers compound, and where documentation
diverges from experience.

The programme has, finally, a policy face. AI-derived accessibility information
circulating through destination platforms needs governance: provenance and time-stamp
standards distinguishing verified fieldwork from automated inference, alignment with the
reporting obligations of the \citet{EAA2019}, and recognition by heritage institutions
that publishing an accessibility claim is an act with consequences rather than a
marketing gesture. Hybrid arrangements in which computation assists experts, managers,
and disabled visitors are where tourism research on AI is heading
\citep{Sigala2024,Tuomi2025}.

Taken together, these developments point to a living, probabilistic accessibility atlas
in which each site's accessibility is a posterior distribution rather than a static score,
updated by Bayesian evidence synthesis as field audits, AI-captured documentation, and
visitor reports accumulate, calibrated across instruments and countries, and displayed
with its uncertainty rather than despite it. Sparse expert fieldwork would then do what
it is uniquely able to do, anchor and calibrate the cheaper evidence streams, while every
new observation sharpens rather than replaces the assessment.

\section{Conclusion}
\label{sec:conclusion}

Accessibility is not a property of places but a relationship between people, heritage
environments, and systems of interpretation, and archaeological sites expose the tensions
between conservation, authenticity, and inclusion more sharply than most tourism
settings.

AI can document, organise, and communicate accessibility information at unprecedented
scale, and our pilot shows it doing useful work there. The same pilot suggests why
evaluation cannot be delegated to systems of this kind: across three runs the desk
instrument measured digital self-presentation and overlooked the embodied and
organisational dimensions that constitute accessibility as lived experience, and even
expert fieldwork supports far less ranking confidence than its averages suggest.
Assessment should therefore use AI for capture and communication and rely on
interpretable inference for evaluation, because inclusion is the domain in which being able to say why a judgement was reached, and how certain it is,
forms part of the accessibility being measured.

\section*{Declaration of generative AI use}
\noindent Generative AI was used in two capacities. First, as the object of study: the
pilot desk audit of Section~\ref{sec:casestudy} was performed by large language models
with web search access under the protocol of Section~\ref{sm:protocol},
and their outputs are reported as data. Second, a large language model was used as a
writing and editing assistant under the direction of the authors, who designed the
study, verified all sources, analyses, and claims, and take full responsibility for the
manuscript.

\section*{Disclosure statement}
\noindent The authors report there are no competing interests to declare.

\section*{Funding}
\noindent No funding was obtained for this work.

\section*{Data availability statement}
\noindent The fieldwork assessment data derive from \citet{Armenio2026}. The desk-audit
protocol, evidence logs, and replication runs are reported in Sections~\ref{sm:protocol}, \ref{sm:log}, and \ref{sm:runs}.

\bibliographystyle{apalike}
\bibliography{references}

\clearpage
\appendix
\setcounter{section}{0}
\setcounter{table}{0}
\setcounter{figure}{0}
\renewcommand{\thesection}{S\arabic{section}}
\renewcommand{\thetable}{S\arabic{table}}
\renewcommand{\thefigure}{S\arabic{figure}}
\renewcommand{\theHsection}{S\arabic{section}}
\renewcommand{\theHtable}{S\arabic{table}}
\renewcommand{\theHfigure}{S\arabic{figure}}

\begin{center}
{\Large\bfseries Supplementary material}
\end{center}
\addcontentsline{toc}{section}{Supplementary material}

\noindent This supplementary material contains the desk-audit protocol
(Section~\ref{sm:protocol}), the first-run desk-audit evidence log
(Section~\ref{sm:log}), the specification of the statistical models
(Section~\ref{sm:models}), and the two replication runs of the desk audit
(Section~\ref{sm:runs}) referred to in the main text.

\section{Desk-audit protocol}
\label{sm:protocol}

The pilot audit reported in Section~\ref{sec:casestudy} was performed on 6 July 2026 by
Claude Fable 5 (Anthropic), accessed through the Claude interface with extended
reasoning enabled and integrated web search as the only information source, in a single
pass; the system was instructed to justify every score from retrieved evidence alone.
For each park the system received the same instruction:

\begin{quote}\small
Using only publicly available online information, assess [park name] on each of the
following seven dimensions of accessibility: physical; sensory; cognitive; inclusive
programmes; staff training; technology and digital; evaluation and monitoring. Score
each dimension on a scale from 1 to 5. Score a dimension only if the evidence you
retrieve supports a judgement; otherwise record it as not assessable. Note the evidence
behind every score.
\end{quote}

Scores were transcribed into Table~\ref{tab:audit} and the supporting evidence into
Section~\ref{sm:log} below; the full transcripts, with the sources retrieved for each
score, are available from the authors on request. Two replication runs under the same
instruction, on a later date and with a second model, are reported in
Section~\ref{sm:runs}.

\section{Desk-audit evidence log}
\label{sm:log}

Condensed from the audit transcripts; scores in parentheses, \na\ marks dimensions
recorded as not assessable for lack of evidence.

\medskip
\noindent\textbf{Cueva Pintada.} Physical (5): self-described fully accessible; lifts,
walkways, step-free routes. Sensory (4): audio-description; tactile visits;
sign-language visits; subtitled audiovisuals. Cognitive (4): comprehensible-content
commitment; autism visits with pictograms. Programmes (4): inclusive educational and
family programmes; inclusive-museum seminar. Staff (3): staff along route; interpreters
on request; no training documented. Technology (4): immersive audiovisual; online
ticketing; digital section. Evaluation (\na): no monitoring trace.

\medskip
\noindent\textbf{Valle dei Templi.} Physical (4): official fully-accessible
designation, free electric wheelchairs, walkways; independent reports of gradients near
ten percent, scarce accessible parking, inaccessible sub-areas. Sensory (4): braille
panels; sign-language video-guide; sensory garden; visits for blind and deaf visitors.
Cognitive (2): little beyond standard panels; route maps. Programmes (3): dedicated
visits; participatory events with disability organisations. Staff (\na): nothing
documented. Technology (4): sign-language application; QR-coded monuments; online
resources. Evaluation (3): publicised participatory assessment initiative.

\medskip
\noindent\textbf{Cerro de la Cabeza.} Physical (2): steep hillside; universal
accessibility only announced in a future plan. Sensory (\na): no evidence. Cognitive
(3): interpretation centre with models and audiovisuals; reconstruction panels.
Programmes (3): school programmes, workshops, guided visits. Staff (\na): no evidence.
Technology (4): on-site virtual-reality headsets; augmented-reality application; online
three-dimensional models. Evaluation (\na): no evidence.

\medskip
\noindent\textbf{Tiermes.} Physical (2): works tendered in 2026 to remedy routes in
poor condition; short seasonal adapted route. Sensory (\na): no evidence. Cognitive
(3): well-regarded interpretive panel network with reconstructions. Programmes (2):
free seasonal guided visits, one adapted. Staff (\na): no evidence. Technology (1): no
application, virtual tour, or digital provision found. Evaluation (2): dated mention of
a continuing visitor survey.

\medskip
\noindent\textbf{Solunto.} Physical (2): hillside setting, visitor advice implying
demanding terrain; no site accessibility information. Sensory, cognitive, programmes,
staff (\na): no evidence. Technology (3): website accessibility widget; announced
QR-code and virtual-tour project. Evaluation (\na): no evidence.

\section{Statistical models}
\label{sm:models}

\noindent\textbf{Field model.} Field scores are modelled as
$F_{ij} = \mu + a_i + b_j + \varepsilon^{F}_{ij}$, with site components
$a_i \sim N(0, \tau_a^2)$, dimension components $b_j \sim N(0, \tau_b^2)$, noise
$\varepsilon^{F}_{ij} \sim N(0, \sigma_F^2)$, an improper flat prior on $\mu$, and
inverse-gamma$(2,1)$ priors on all variance components. Site-level accessibility is
$\theta_i = \mu + a_i$.

\medskip
\noindent\textbf{Calibration model.} Desk scores, where observed, are modelled as
$D_{ij} = \mu + a_i + b_j + \delta_0 + \delta_j + \varepsilon^{D}_{ij}$, sharing
$\mu, a_i, b_j$ with the field model, with overall desk shift $\delta_0 \sim N(0, 1)$,
dimension-specific shifts $\delta_j \sim N(0, \tau_\delta^2)$, desk noise
$\varepsilon^{D}_{ij} \sim N(0, \sigma_D^2)$, and inverse-gamma$(2,1)$ priors on
$\tau_\delta^2$ and $\sigma_D^2$. The reported desk shift per dimension is
$\Delta_j = \delta_0 + \delta_j$. Desk missingness is conditioned on, not modelled; the
blanks are informative and a full treatment would model them.

\medskip
\noindent\textbf{Computation and summaries.} Both models are fitted by conjugate Gibbs
sampling (24{,}000 iterations, 6{,}000 burn-in, fixed seed); chains mix rapidly and
results are stable across seeds. Posterior means and 90\% intervals are reported below;
the site levels $\theta_i$ are those displayed in Figure~\ref{fig:posterior}. Dimension
components $b_j$: physical $+0.50$ $(-0.05, 1.05)$; sensory $+0.33$ $(-0.22, 0.89)$;
cognitive $-0.36$ $(-0.91, 0.19)$; programmes $+0.33$ $(-0.21, 0.88)$; staff $-0.18$
$(-0.74, 0.36)$; technology $+0.16$ $(-0.39, 0.70)$; evaluation $-0.87$
$(-1.45, -0.32)$. Desk shifts $\Delta_j$: physical $+0.01$ $(-0.58, 0.58)$; sensory
$+0.32$ $(-0.41, 1.06)$; cognitive $+0.46$ $(-0.15, 1.08)$; programmes $+0.04$
$(-0.58, 0.65)$; staff $+0.01$ $(-0.86, 0.85)$; technology $+0.37$ $(-0.21, 0.95)$;
evaluation $+0.54$ $(-0.20, 1.30)$. Noise: $\sigma_F$ 0.56 $(0.45, 0.69)$; $\sigma_D$
0.77 $(0.59, 0.99)$. Under the calibration model the 90\% intervals for $\theta_i$
narrow from an average width of 1.08 to 0.99. The sensory, staff, and evaluation shifts
rest on one or two jointly observed parks and should be read accordingly.

\section{Replication runs of the desk audit}
\label{sm:runs}

\noindent\textbf{Design.} The audit of Section~\ref{sm:protocol} was repeated on
18 September 2026 with two models, Claude Fable 5.1 and Claude Opus 5 (Anthropic), each
with integrated web search. Each park was audited in a separate, fresh session with no
conversation history and no access to the field scores, to Table~\ref{tab:audit}, or to the first-run
log of Section~\ref{sm:log}; ten sessions in total. Each session received the
instruction of Section~\ref{sm:protocol} verbatim, preceded by a procedural wrapper: use
only publicly available online information retrieved during the session and cite every
score by URL; search in English and in the local language (Spanish or Italian) with
between six and fifteen searches; draw on the site's official pages and social channels,
public institutional documents, accessibility portals, news, and visitor reviews;
exclude academic theses, articles, or preprints reporting accessibility assessments of
the park; do not force a score, since ``not assessable'' is an expected outcome; and note
the date of evidence, distinguishing announced works from existing provision. Sessions
returned scores, one-line evidence summaries with URLs, the list of queries run, and any
conflicts among sources. On the audit date several official pages could not be
retrieved by the search tool (the Cueva Pintada accessibility page, the Valle dei Templi
park website including its accessibility document and service charter, the Cerro de la
Cabeza park website, and the Tiermes museum site), which the sessions recorded as
caveats; scores rest on institutional, press, portal, and review sources.

\medskip
\noindent\textbf{Scores.} Table~\ref{tab:runs} reports, for every park and dimension,
the field score and the three desk readings. Table~\ref{tab:agree} summarises coverage
and agreement.

\begin{table}[h]
\centering
\caption{Field score and desk readings across the three runs, as F / R1 / R2 / R3.
F: fieldwork \citep{Armenio2026}. R1: Claude Fable 5, 6 July 2026
(Table~\ref{tab:audit}). R2: Claude Fable 5.1, 18 September 2026. R3: Claude Opus 5, 18 September 2026.
\na\ = recorded as not assessable.}
\label{tab:runs}
\footnotesize
\setlength{\tabcolsep}{3.5pt}
\begin{tabular}{lccccccc}
\toprule
Park & Phys. & Sens. & Cogn. & Progr. & Staff & Tech. & Eval. \\
\midrule
Cueva Pintada      & 4/5/4/5 & 4/4/2/2 & 3/4/3/3 & 4/4/4/3 & 3/3/\na/\na & 4/4/3/3 & 3/\na/2/2 \\
Valle dei Templi   & 4/4/4/4 & 4/4/4/4 & 2/2/2/1 & 3/3/2/3 & 3/\na/\na/2 & 4/4/3/3 & 1/3/2/2 \\
Cerro de la Cabeza & 2/2/2/2 & 3/\na/2/2 & 2/3/2/2 & 3/3/2/\na & 2/\na/\na/\na & 2/4/3/3 & 2/\na/\na/\na \\
Tiermes            & 3/2/2/3 & 2/\na/\na/1 & 2/3/\na/1 & 3/2/\na/2 & 2/\na/\na/\na & 2/1/1/2 & 1/2/\na/\na \\
Solunto            & 3/2/1/1 & 2/\na/\na/\na & 2/\na/2/2 & 2/\na/1/2 & 2/\na/\na/\na & 2/3/2/3 & 1/\na/\na/\na \\
\bottomrule
\end{tabular}
\end{table}

\begin{table}[h]
\centering
\caption{Coverage of each run and pairwise agreement between runs. Status agreement:
share of the 35 cells with the same assessable or not-assessable status in both runs.
Jointly scored: cells scored in both runs. Exact and within-one: share of jointly scored
cells with identical scores, and with scores differing by at most one point. MAD:
mean absolute deviation between the two runs' scores (between runs), or between a run
and the field scores (versus field).}
\label{tab:agree}
\footnotesize
\setlength{\tabcolsep}{4pt}
\begin{tabular}{lcccc}
\toprule
Run & Coverage & MAD vs field & Mean signed diff. & Over / equal / under \\
\midrule
R1 (Fable 5, July)      & 23/35 & 0.61 & $+0.26$ & 8 / 11 / 4 \\
R2 (Fable 5.1, Sept.)   & 23/35 & 0.65 & $-0.48$ & 2 / 10 / 11 \\
R3 (Opus 5, Sept.)      & 26/35 & 0.69 & $-0.38$ & 4 / 10 / 12 \\
\midrule
Pair & Status agreement & Jointly scored & Exact / within one & MAD between runs \\
\midrule
R1 vs R2 & 27/35 (77\%) & 19 & 37\% / 95\% & 0.68 \\
R1 vs R3 & 26/35 (74\%) & 20 & 35\% / 90\% & 0.75 \\
R2 vs R3 & 30/35 (86\%) & 22 & 64\% / 100\% & 0.36 \\
\bottomrule
\end{tabular}
\end{table}

\noindent Per dimension, staff training was scored in one park in R1, none in R2, and
one in R3; evaluation and monitoring in two parks in every run; sensory provision in
two, three, and four parks. Six cells were not assessable in any run: staff training at
Cerro de la Cabeza, Tiermes, and Solunto; evaluation at Cerro de la Cabeza and Solunto;
sensory provision at Solunto. Twenty-four of the 35 cells had the same status in all
three runs. The two September runs, which shared a date, agreed with each other more
closely than either did with July, consistent with part of the between-run variation
being driven by what was retrievable on the day rather than by the model. The first-run
sign pattern (desk above field where announcements dominate) reversed on average in
September, when the official pages of the two best-documented parks were unreachable
and scores fell back on third-party audits, press, and reviews; the three cells singled
out in Section~\ref{sec:pilot} nevertheless stayed above the field in every run
(Cueva Pintada physical: field 4, desk 5/4/5; Valle dei Templi evaluation: field 1,
desk 3/2/2; Cerro de la Cabeza technology: field 2, desk 4/3/3).

\medskip
\noindent\textbf{Condensed evidence logs, runs 2 and 3.} Scores in parentheses; sources
are summarised, with full URLs in the session outputs held by the authors.

\medskip
\noindent\textbf{Cueva Pintada.} \emph{R2:} Physical (4): TUR4all audit sheet (undated)
and PREDIF recognition (2010) record a step-free chain, lift with braille and audio
buttons, adapted toilet, lowered desk; not 5 for lack of a dated independent
confirmation. Sensory (2): braille and contrast signage per TUR4all; 2021 press on QR
codes and sign-language work ``being done''; no audio description, loop, or tactile
models found. Cognitive (3): recurring autism-oriented visits with a pictogram document
(2024--2025); no easy-read material for all visitors. Programmes (4): three dated
autism family workshops, adapted visits listed by TUR4all, 2021 inclusive-museum
seminar. Staff (\na): seminar participation only; no training documented. Technology
(3): virtual visit, 3D collection, museums app, 2022 VR exhibition; no accessibility
claims for the digital offer. Evaluation (2): director reports continuous visitor
studies with a university partner since 2006; annual visitor statistics; no
accessibility-specific monitoring. \emph{R3:} Physical (5): the same two audits read as
a complete step-free chain. Sensory (2): building-level features only. Cognitive (3):
pictogram document for autistic visitors, repeated workshops. Programmes (3): adapted
visits and autism workshops; no provision evidenced for other groups. Staff (\na):
island-wide accessibility plan proposes training but does not name the site.
Technology (3): 3D and audiovisual displays, multilingual tours, virtual visit, museums
app with a Cabildo accessibility statement. Evaluation (2): a 2007--2009 visitor
satisfaction study; nothing accessibility-specific or recent. Both runs note that the
museum's own accessibility page was unreachable.

\medskip
\noindent\textbf{Valle dei Templi.} \emph{R2:} Physical (4): 85\% of the site declared
barrier-free after 2020 works, free electric wheelchairs and shuttle, accessible
toilets; conflicting wheelchair counts (two on the concessionaire's page, six in the
regional release) and a 2025 review calling close access to the temples very
difficult. Sensory (4): braille panels and tactile map, tactile Telamon, sign-language
videos by QR code, sensory garden (2023), VideoSegnArt app (2024). Cognitive (2): no
park-run easy-read or social-story material; the only such provision is in the
FAI-run Kolymbethra garden. Programmes (2): episodic events on the International Day of
Persons with Disabilities and an autism walk hosted rather than designed by the park.
Staff (\na): 2019 announcement that sign-language training was initiated, no later
confirmation. Technology (3): VideoSegnArt app, QR sign-language panels, audioguide;
website conformance unverifiable. Evaluation (2): a 2019 participatory needs
assessment with eight disability associations and the director's own testing; no
published audit or survey. \emph{R3:} Physical (4): as R2, with visitor reports of
steps at the Temple of Juno and no buggy cover in the western sector. Sensory (4): as
R2. Cognitive (1): no cognitive-specific provision found; a 2019 source names it as
still needed. Programmes (3): standing concessionaire offer of multisensory visits plus
recurring autism-day events with park patronage. Staff (2): the 2019 sign-language
training for custodial staff, reported as under way, with no later evidence.
Technology (3): as R2. Evaluation (2): as R2; the service charter page could not be
retrieved.

\medskip
\noindent\textbf{Cerro de la Cabeza.} \emph{R2:} Physical (2): a 2019 third-party audit
of the interpretation centre (17\% kerb ramp needing assistance, 4~cm entrance step,
no reserved bays, two adapted toilets); the hilltop site has no documented accessible
route and is temporarily closed for works whose stated goal is an accessible itinerary.
Sensory (2): high-contrast panels and toilet pictograms, low lighting; no braille,
tactile, audio, loop, or sign-language provision. Cognitive (2): panels, scale models,
audiovisuals, VR and AR reconstructions; nothing easy-read; cognitive accessibility is
a future project aim. Programmes (2): generic guided visits, workshops, and school
visits; nothing for disabled visitors. Staff (\na): no evidence. Technology (3): VR
sessions and the regional AR app since 2024, rated 3.2 with signal complaints; no
accessibility features documented. Evaluation (\na): no surveys, statistics, or
monitoring found. \emph{R3:} Physical (2): as R2. Sensory (2): as R2, plus a regional
statement that the XR experiences are adapted for hearing loss. Cognitive (2): as R2,
with visitor reports of poor signage. Programmes (\na): school and workshop provision
only; nothing targeted found. Staff (\na): a 2014 tourism-quality award with generic
awareness sessions; nothing disability-specific. Technology (3): as R2; municipal
tourism site claims an obsolete web-accessibility standard. Evaluation (\na): as R2.
Both runs note a funded but unexecuted project promising universal accessibility.

\medskip
\noindent\textbf{Tiermes.} \emph{R2:} Physical (2): regional works completed August
2026 (regravelled paths, restored walkway, extra steps) improve comfort rather than
step-free access; guides and reviews state the site is unsuitable for wheelchairs.
Sensory (\na): no provision mentioned anywhere. Cognitive (\na): standard panels and a
numbered map only; reviews report broken signs and an unclear route. Programmes (\na):
generic guided tours; museum education page empty. Staff (\na): no evidence.
Technology (1): no app, audioguide, QR codes, or virtual tour; only the regional
museums portal's accessibility features. Evaluation (\na): regional aggregate
statistics only. \emph{R3:} Physical (3): the same 2026 works, plus a 45-minute
reduced-mobility circuit documented in 2010 and 2017 sources while the regional
directory lists wheelchair access as limited. Sensory (1): service listings enumerate
other facilities and nothing sensory. Cognitive (1): no easy-read or structured-visit
material; wayfinding criticised. Programmes (2): free summer guided visits,
archaeologist-led tours, astronomy events; free entry for visitors with a disability
certificate; nothing designed for disabled audiences. Staff (\na): regional training
references only, not site-specific. Technology (2): accessible regional portal; an iOS
app listing found but not retrievable; no on-site digital provision. Evaluation
(\na): a 2010 announcement of visitor surveys; nothing since.

\medskip
\noindent\textbf{Solunto.} \emph{R2:} Physical (1): steep unpaved hillside road,
reviews advising against visiting with walking difficulty, no official accessibility
information, no shuttle or mobility aid. Sensory (\na): the only sensory aid in the
park cluster (sign-language smartglasses) is at Himera. Cognitive (2): bilingual
boards and a photocopied map; conflicting reviews on signage quality. Programmes (1):
general tours, free days, and an invitation-only event; nothing targeted. Staff (\na):
review praise for individual staff only. Technology (2): the park cluster's app,
virtual tours, and AR are documented for Himera, not Solunto; website accessibility
widget only. Evaluation (\na): no survey, charter, or audit found. \emph{R3:} Physical
(1): as R2. Sensory (\na): as R2. Cognitive (2): thin interpretation plus an
illustrated guide for young readers (2025). Programmes (2): active school and public
programme including Solunto; free entry for disabled visitors and a companion; nothing
co-designed with disabled audiences. Staff (\na): as R2. Technology (3): park app,
inclusion in a regional digitisation circuit (2022), a necropolis virtual tour reported
as forthcoming, website accessibility widget. Evaluation (\na): as R2.

\end{document}